\documentclass[letterpaper,twocolumn,10pt]{article}
\usepackage{usenix2020-09}
\usepackage{amssymb} 
\usepackage{tikz}
\usepackage{amsmath}
\usepackage{url}

\newcommand{\sysname}{TetriInfer}

\newcommand{\LPLD}{\textsc{LightPre-LightDec}}
\newcommand{\LPHD}{\textsc{LightPre-HeavyDec}}
\newcommand{\HPLD}{\textsc{HeavyPre-LightDec}}
\newcommand{\HPHD}{\textsc{HeavyPre-HeavyDec}}

\newcommand{\LP}{light prefill}
\newcommand{\HP}{heavy prefill}
\newcommand{\HD}{heavy decode}
\newcommand{\LD}{light decode}

\newcommand{\cc}[1]{{{\color{blue}#1}}{}}

\usepackage{tikz}

\begin{document}

\date{}


\title{Inference without Interference: \\ Disaggregate LLM Inference for Mixed Downstream Workloads}

\author{
\rm
Cunchen Hu\textsuperscript{1,2}\footnotemark{},
Heyang Huang\textsuperscript{1,2},
Liangliang Xu\textsuperscript{3},
Xusheng Chen\textsuperscript{3},
Jiang Xu\textsuperscript{3},
Shuang Chen\textsuperscript{3}, \\
\rm Hao Feng\textsuperscript{3},
Chenxi Wang\textsuperscript{1,2},
Sa Wang\textsuperscript{1,2},
Yungang Bao\textsuperscript{1,2},
Ninghui Sun\textsuperscript{1,2},
Yizhou Shan\textsuperscript{3} \\
\break \\
\rm \large\textsuperscript{1}\textit{University of Chinese Academy of Sciences}, \textsuperscript{2}\textit{ICT, CAS}
\textsuperscript{3}\textit{Huawei Cloud}}

\maketitle

\footnotetext{Work done while intern at Huawei Cloud.}

\begin{abstract}

Transformer-based large language model (LLM) inference serving is now the backbone of many cloud services.
LLM inference consists of a prefill phase and a decode phase.
However, existing LLM deployment practices often overlook the distinct characteristics of these phases, leading to significant interference.
To mitigate interference, our insight is to carefully schedule and group inference requests based on their characteristics. We realize this idea in \sysname\ through three pillars. First, it partitions prompts into fixed-size chunks so that the accelerator always runs close to its computation-saturated limit. Second, it disaggregates prefill and decode instances so each can run independently. Finally, it uses a smart two-level scheduling algorithm augmented with predicted resource usage to avoid decode scheduling hotspots.
Results show that \sysname\ improves time-to-first-token (TTFT), job completion time (JCT), and inference efficiency in turns of performance per dollar by a large margin, e.g., it uses 38\% less resources all the while lowering average TTFT and average JCT by 97\% and 47\%, respectively.

\end{abstract}
\section{Introduction}

Since the boom of ChatGPT, large language model (LLM) based services have now played a vital role in our daily lives\cite{evaluating-arvix21, llama-arvix23, promise-nature23, towards-sigkdd23, zero-icml21, memgpt-arvix23}.
Behind the scenes, all use cases boil down to LLM inference serving. To run an inference request, the LLM model will first take the user inputs to generate the first token (known as the prefill phase), and then generate outputs token-by-token in an auto-regressive manner (known as the decode phase).
Numerous works were proposed to improve the cost efficiency of LLM inference~\cite{vllm-sosp23, fastserve-arxiv23}.

There are various ways to interact with LLM, from simple chats to more complex downstream tasks such as document summarization, content creation, etc.
As a result, LLM-empowered services serve inference requests with dramatically different properties that can be categorized across two dimensions: the input prompt length during the prefill phase and the generated token length during the decode phase.
As shown in Figure~\ref{fig-dataset-cdf}, summarization tasks have long input prompts and short generated tokens, while context creation tasks are the opposite. 
Token lengths of different downstream tasks can differ by more than two orders of magnitude.
Given the significant variation in LLM inference requests from various downstream tasks, the first research question we ask in this paper is \textit{how do these inference requests perform when running together?}. 

To answer this question, we run extensive tests that mix LLM prefill and decode requests of different lengths. 
Unfortunately, we have observed serious interference across all combinations. For example, mixing prefill requests could result in a 10x slowdown, combining prefill and decode requests could lead to a 5x slowdown, and mixing decode requests with different lengths could take a 16\% throughput hit (see \S\ref{sec-bg-motivation}). 
A naive solution to avoid interference is to provision resources for each downstream task statically. Given the high cost of LLM serving infrastructure, this solution is impractical.
To this end, the second research question we ask in this paper is \textit{how to build a distributed LLM inference serving system that minimizes interferences?}

We take a step back to examine why interference exists. We find the fundamental issue lies in the fact that current LLM deployment practices do not account for the distinct characteristics exhibited by LLM prefill and decode phases.
Specifically, the prefill phase resembles a computation-heavy batch job, with its computation scaling quadratically with the input prompt length.
The decode phase resembles a memory-intensive, latency-critical task, with its resource usage scaling sublinearly with the generated token length~\cite{pope2023efficiently}.
Interferences observed in our tests are classic system problems.
Running prefill requests leads to a serious slowdown because we continue adding computation-heavy jobs to an already saturated hardware (\S\ref{sec-bg-pp}).
Combining prefill and decode requests hurts both because we co-run batch and latency-critical jobs simultaneously (\S\ref{sec-bg-pd}).
Mixing decode requests leads to a throughput drop because we are unaware of the memory bandwidth and capacity usage, thus leading to contention and head-of-line blocking (\S\ref{sec-bg-dd}).

To solve these issues, our insight is to carefully \textit{schedule and group requests based on their characteristics}.
We realize this idea in \sysname\footnote{The name of our system, \sysname, implies that it can efficiently organize LLM inference requests, similar to how tetris blocks are stacked.}, a cloud-scale LLM inference
serving system designed to battle interferences.

Our designs are three-fold.
First, to avoid interference running prefill, we propose limiting the number of tokens processed in a single prefill iteration so that hardware is fully utilized without incurring extra penalties. \sysname\ partitions and pads input prompts into fixed-size chunks so that the accelerator always runs close to its computation-saturated limit (\S\ref{sec-prefill-instance}).
Second, to avoid interference in co-running prefill and decode, we propose disaggregating prefill from decode phases.
\sysname\ has dedicated prefill and decode instances.
During runtime, prefill instances transfer prefilled KV cache to decode instances.
The prefill and decode instances are virtual concepts in that
each can scale independently and flip roles if load changes (\S\ref{sec-instance-transition}).
Third, to avoid interference running decode requests, we propose using a smart two-level scheduling algorithm augmented with predicted resource usage to avoid scheduling hotspots (\S\ref{sec-decode-instance}). \sysname\ incorporates an LLM-based length prediction model to speculate the number of generated tokens of decode requests, and then schedule them accordingly.

We implement \sysname's disaggregated prefill and decode instances based on vLLM~\cite{vllm-sosp23}. Most of our modules are implemented in Python, except for the network stack module, which utilizes C++ to interface with low-level APIs for KV cache transfer. The fine-tuning part uses Trainer APIs offered by HuggingFace Transformer~\cite{url-hg-opt125}. Since we cannot access high-end hardware, we implement a mock mechanism to emulate varying network bandwidth connecting prefill and decode instances, as illustrated in Figure~\ref{fig-network-stack}.

We compare \sysname\ with vanilla vLLM using public dataset~\cite{sharegpt} in terms of time-to-first-token (TTFT), job completion time (JCT), and efficiency as in performance per dollar (perf/\$).
We run them atop a real testbed with emulated network bandwidth ranging from 200Gbps to 300GBps.
For light prefill and heavy decode workload, \sysname\ improves perf/\$ by 2.4x (Figure~\ref{fig-ben-lphd-dataset}). For common mixed workload, \sysname\ improves average TTFT and average JCT by 85\% and 50\%, respectively (Figure~\ref{fig-ben-mix-dataset}).
Nevertheless, we also find that \sysname's design is not ideal for heavy prefill and heavy decode workloads since the room for improvement is marginal, and the overhead we introduce cannot be offset (Figure~\ref{fig-ben-hphd-dataset}).
Overall, our ideas mentioned above are effective.
\sysname\ achieves effective LLM inference serving, outperforming vLLM by a large margin in TTFT, JCT, and perf/\$ running most common workloads (\S\ref{sec-app-cost-perf}). 

{
\begin{figure}[t]
\begin{center}
\vspace{5pt}
\centerline{\includegraphics[width=0.48\textwidth]{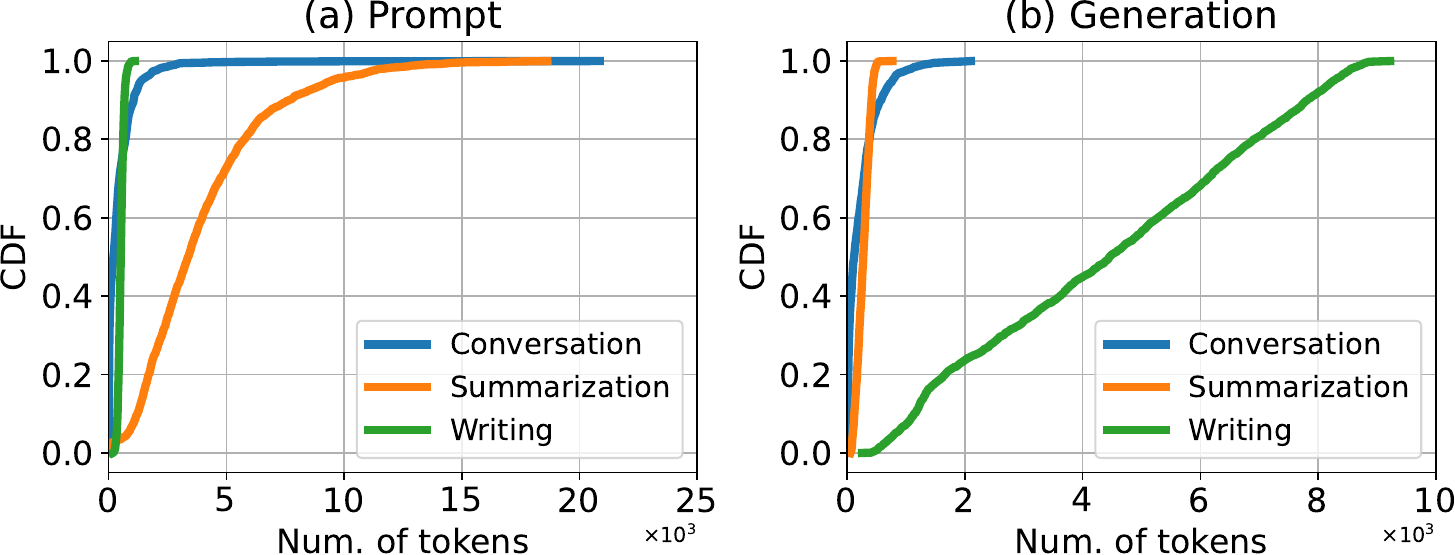}}
\caption{\textbf{Length Distribution.} Prompt Tokens for Prefill and Generated Tokens during Decode. Data sources: conversation~\cite{sharegpt}, summarization~\cite{url-hg-summarization}, writing~\cite{url-hg-writing}.}
\label{fig-dataset-cdf}
\end{center}
\end{figure}
}

\section{Background and Motivation}

We present a brief primer on LLM inference and study interferences while running various LLM inference requests to motivate our work. For model and testbed details, see \S\ref{sec-eval}.

\subsection{Generative LLM Inference}

{
\begin{figure}[t]
\begin{center}
\centerline{\includegraphics[width=0.48\textwidth]{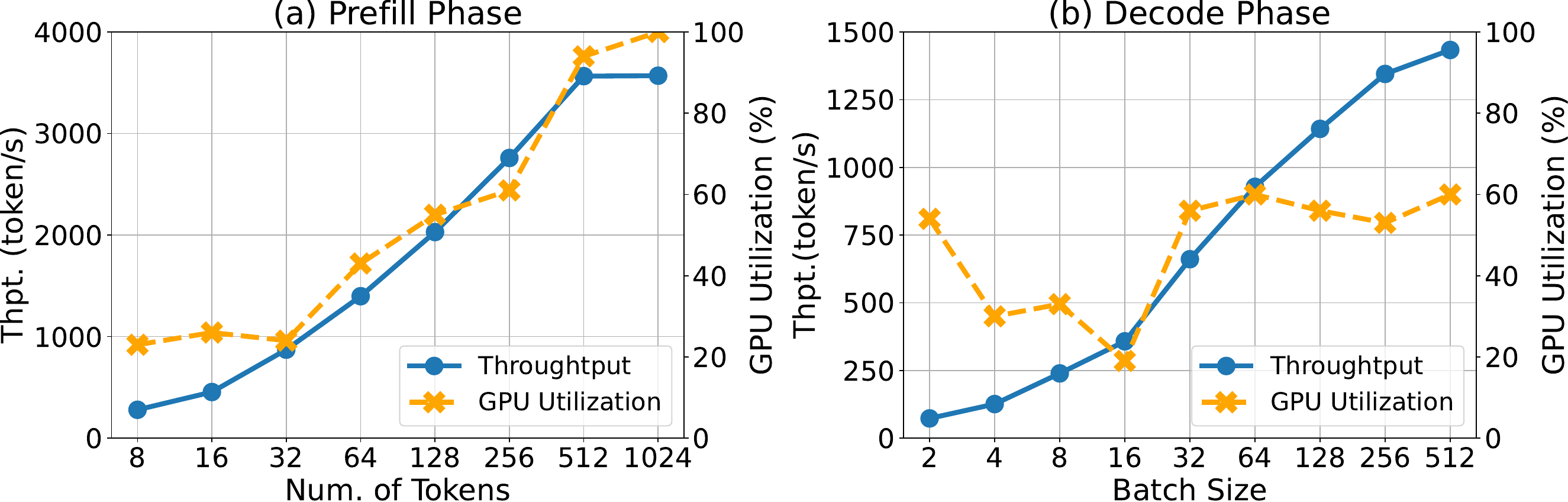}}
\caption{\textbf{Prefill and Decode's Characteristics.} Decode's GPU utilization fluctuates because the task is faster than our monitoring granularity.}
\label{fig-prefill-decode}
\vspace{-0.3in}
\end{center}
\end{figure}
}

LLM inference is a process that involves generating a sequence of output tokens in response to an input prompt. This process consists of two phases: prefill and decode.
The prefill phase outputs the first token and generates the key and value cache (KV cache) for future decoding~\cite{vllm-sosp23}. The decode phase uses the previous KV cache to generate new tokens step-by-step in an auto-regressive manner.
Generally, the prefill phase is computation-bound, and the decode phase is memory-bound~\cite{pope2023efficiently}.
We report this in Figure~\ref{fig-prefill-decode}.
Results indicate that the prefill phase's throughput stays flat once the accelerator is saturated at a certain number of tokens (which we name the accelerator-saturate threshold).
The decode phase's throughput continues increasing with a larger batch size but plateaus once the memory bandwidth is saturated. 

\subsection{Motivation: Interference Study}
\label{sec-bg-motivation}


This section studies the impact of running different inference requests concurrently.
Inspired by Figure~\ref{fig-dataset-cdf},
we classify inference requests across two dimensions (prefill and decode length) and one property (light or heavy), resulting in four distinct request types:
heavy prefill,
light prefill,
heavy decode, and
light decode.
Here, heavy refers to a long token length, while light refers to a short token length.
%
Below, we study mixing prefill and prefill (\S\ref{sec-bg-pp}),  prefill and decode (\S\ref{sec-bg-pd}), decode and decode (\S\ref{sec-bg-dd}). 

\subsubsection{Prefill and Prefill}
\label{sec-bg-pp}

{
\begin{figure}[t]
\begin{center}
\vspace{5pt}
\centerline{\includegraphics[width=0.46\textwidth]{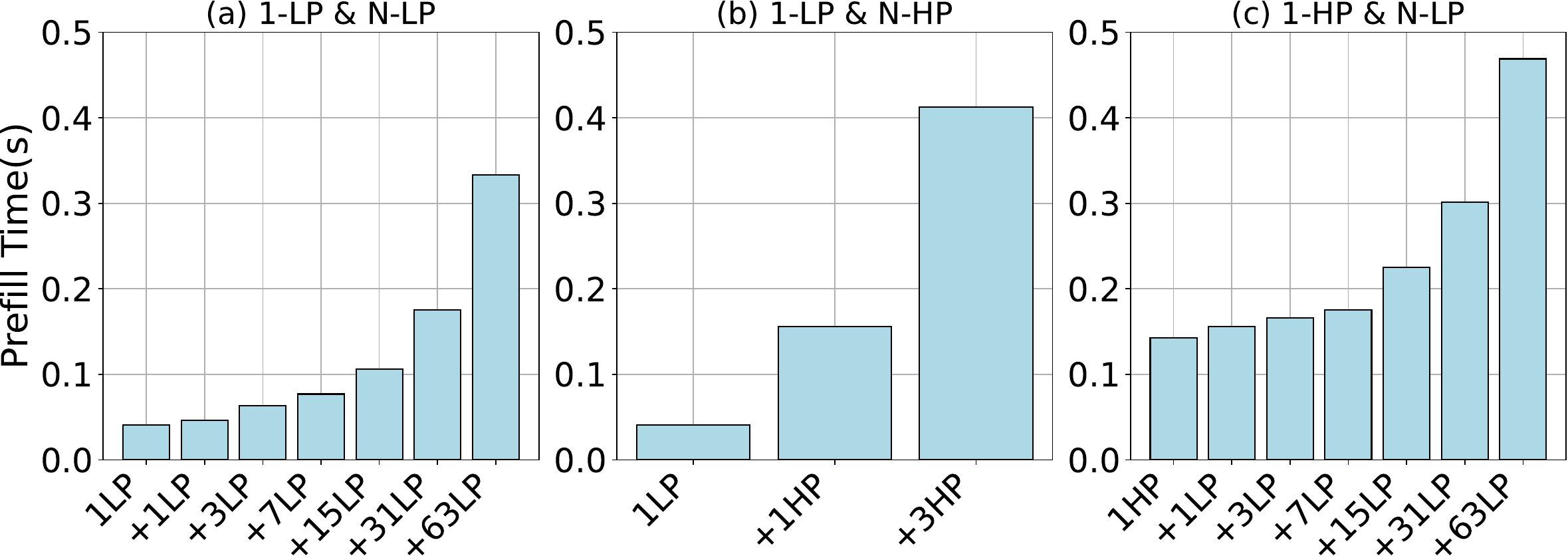}}
\caption{\textbf{Interference of Prefill \& Prefill.} LP means \LP. HP means \HP. (a) and (b) indicate that \LP's prefill latency increases as the number of co-running requests increases. The same applies to (c)'s \HP.}
\vspace{-0.2in}
\label{fig-study-prefill-prefill}
\end{center}
\end{figure}
}

We first study mixing prefill requests.
Here, \LP\ has roughly 18 prompt tokens as it is the median token length in ShareGPT's short prompts~\cite{sharegpt}, while \HP\ has 512 prompt tokens as the accelerator is saturated at this length in our testbed (see Figure~\ref{fig-prefill-decode}).
In Figure~\ref{fig-study-prefill-prefill} (a) and (b), we show how a  \LP's latency changes if it co-runs with other \LP\ and \HP\ requests. We find its latency increases by 2x and 8x if there are 7 and 63 concurrent \LP\ requests in the same batch. Additionally, it incurs more than 10x latency slowdown if it runs with other \HP\ requests.
In Figure~\ref{fig-study-prefill-prefill} (c), we show that \HP's latency also incurs a 3x slowdown if co-run with other \LP\ requests.
Overall, we find that when the total number of tokens in a batch is larger than the accelerator-saturate threshold, the prefill latency dramatically increases.

\subsubsection{Prefill and Decode}
\label{sec-bg-pd}

{
\begin{figure}[t]
\vspace{5pt}
\begin{center}
    \setlength{\abovecaptionskip}{1pt plus 1.0pt minus 1.0pt}
    \setlength{\belowcaptionskip}{1pt plus 1.0pt minus 1.0pt}
\centerline{\includegraphics[width=0.48\textwidth]{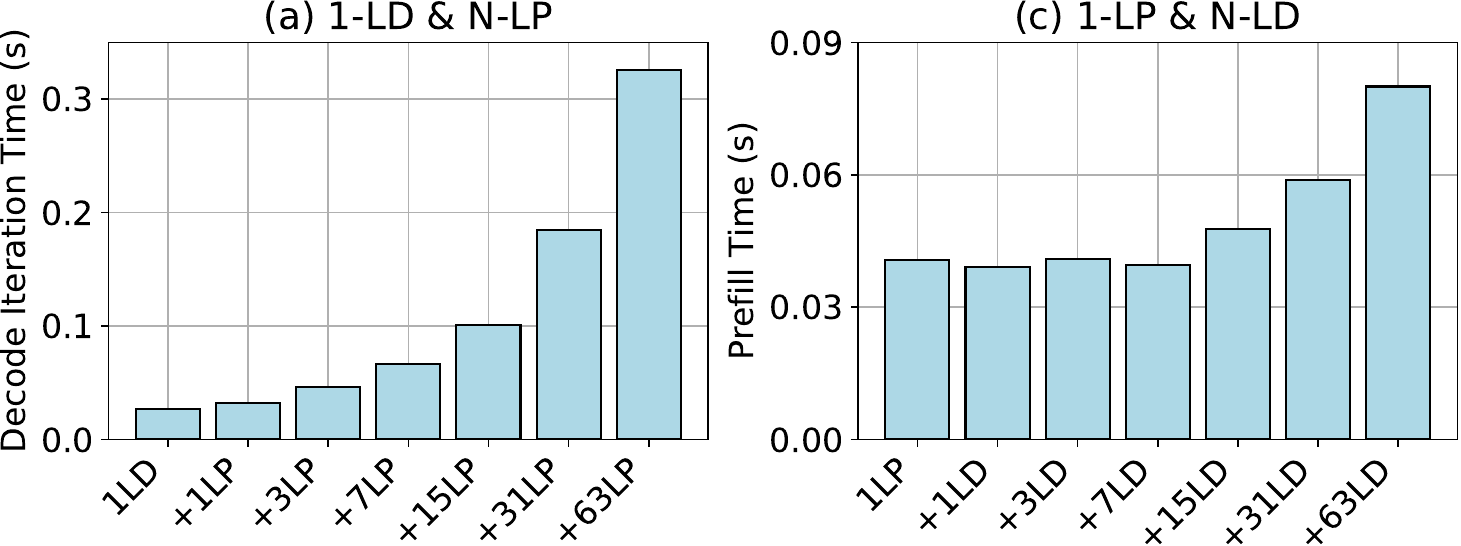}}

\centerline{\includegraphics[width=0.48\textwidth]{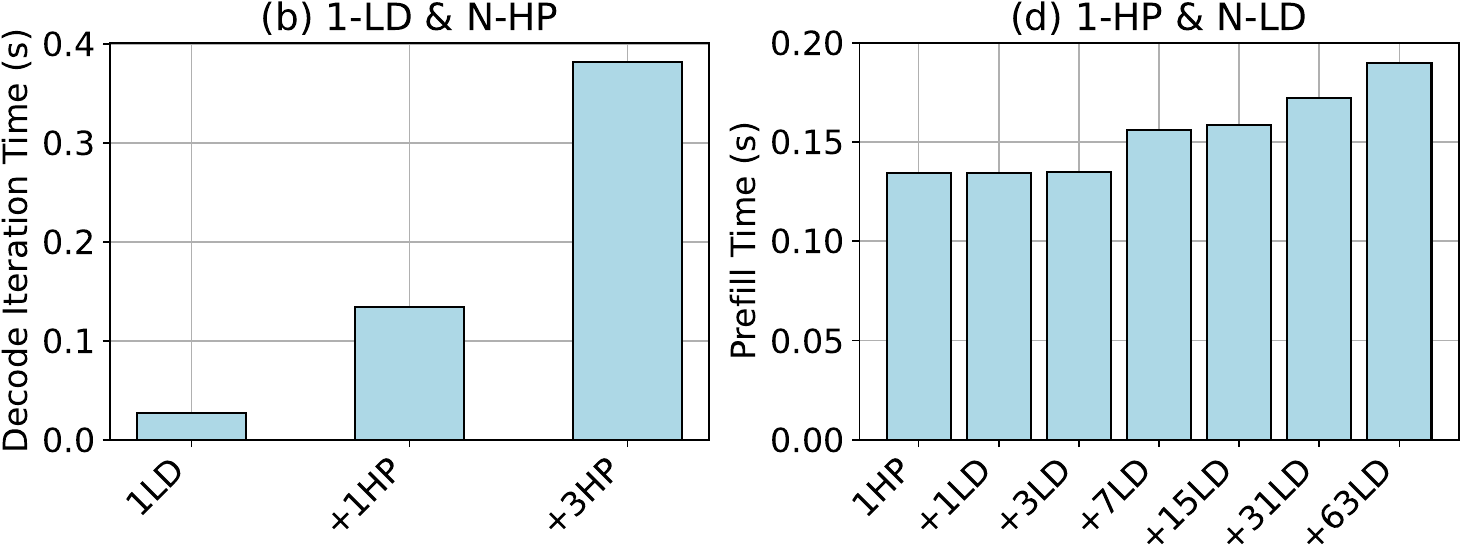}}
\caption{\textbf{Interference of Prefill \& Decode.} LD means \LD. HD means \HD.}
\label{fig-study-prefill-decode}
\end{center}
\end{figure}
}

We now study mixing prefill and decode in the same batch due to continuous batching~\cite{vllm-sosp23,yu2022orca}.
Both \LP\ and \HP\ follow \S\ref{sec-bg-pp}'s definition.
Also, \LD\ refers to the ones that generate a small number of tokens, e.g., less than 100. \HD\ refers to the ones that generate a large number of tokens, e.g., larger than 512 tokens.
Though decoding latency increases slowly with an increasing number of generated tokens, we only present tests related to \LD\ as \HD\ presents similar results.

In Figure~\ref{fig-study-prefill-decode} (a) and (b), we show how a \LD's per-iteration decoding latency changes if it co-runs with \LP\ and \HP\ requests. Results indicate that its decoding latency increases by 5x even if only one other heavy prefill request is in the same continuous batch!
In Figure~\ref{fig-study-prefill-decode} (c) and (d), we show how a \LP's and a \HP's latency changes if they co-run with other \LD\ requests. Figure~\ref{fig-study-prefill-decode} (c) indicates that the prefill latency increases once the number of co-running \LD\ requests is more than 7. Both slow down roughly by up to 2.5x.

\subsubsection{Decode and Decode}
\label{sec-bg-dd}

{
\begin{figure}[t]
\begin{center}
    \setlength{\abovecaptionskip}{3pt plus 1.0pt minus 1.0pt}
    \setlength{\belowcaptionskip}{6pt plus 1.0pt minus 1.0pt}
\centerline{\includegraphics[width=0.48\textwidth]{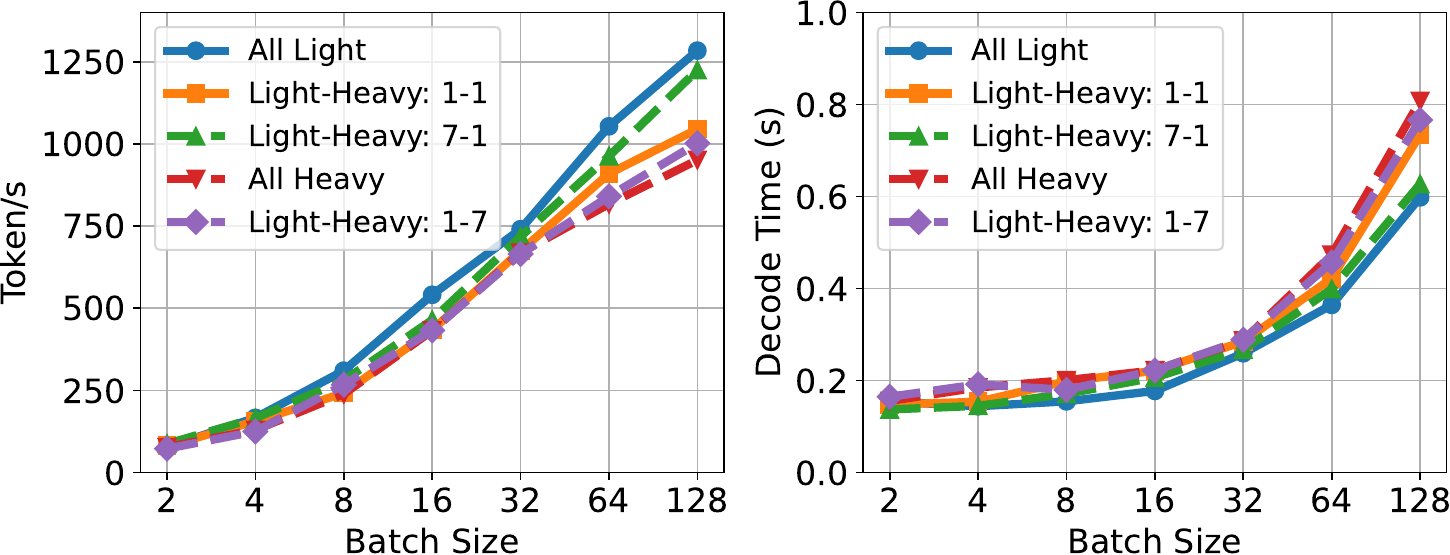}}
\caption{\textbf{Interference of Decode \& Decode.}}
\vspace{-0.3in}
\label{fig-study-decode-decode}
\end{center}
\end{figure}
}

We now study mixing decode requests.
Following \S\ref{sec-bg-pd}'s definition and ShareGPT's distribution, we set both requests to use very short prompts. And \LD\ generates roughly 20 to 100 tokens, while \HD\ generates more than 512 tokens.
Figure~\ref{fig-study-decode-decode} presents the decoding throughput and latency while running different numbers of \LD\ and \HD\ requests in the same batch. Results suggest that compared to a batch with all \LD\ requests, increasing \HD\ requests could seriously hurt throughput and latency. For example, with a batch size of 128, compared to a batch with all \LD, a batch with half \HD\ and half \LD's throughput drops by 16\% while the latency increases by 23\%.

\subsection{Analysis and Insights}

We have observed significant interferences in LLM inferencing.
The root cause is simple: current LLM systems are ignorant of the distinct characteristics exhibited by LLM prefill and decode phases. The prefill phase resembles a computation-heavy batch job, while the decode phase resembles a memory-intensive, latency-critical task~\cite{pope2023efficiently}.

Interferences measured above are classic system problems.
In \S\ref{sec-bg-pp}, running prefill requests leads to a serious slowdown because we continue adding computation-heavy jobs to an already saturated hardware.
In \S\ref{sec-bg-pd}, mixing prefill and decode requests hurts both because we co-run batch and latency-critical jobs at the same time.
In \S\ref{sec-bg-dd}, mixing decode requests leads to a throughput drop because we are unaware of the memory bandwidth and capacity usage, thus leading to contention and head-of-line blocking.

Our work aims to solve these issues by carefully \textit{schedule and group requests based on their characteristics}.
Our ideas are three-fold.
First, to avoid interference running prefill, we propose limiting the number of tokens processed in a single prefill step so that hardware is fully utilized without incurring extra penalties.
Second, to avoid interference co-running prefill and decode, we propose disaggregating prefill from decode so that each runs independently.
Third, to avoid interference running decode requests, we propose to use a smart two-level scheduling algorithm augmented with predicted resource usage to avoid scheduling hotspots. We visualize the comparison in Figure~\ref{fig-architecture} (a).

{
\begin{figure*}[ht]
\begin{center}
\centerline{\includegraphics[width=0.95\textwidth]{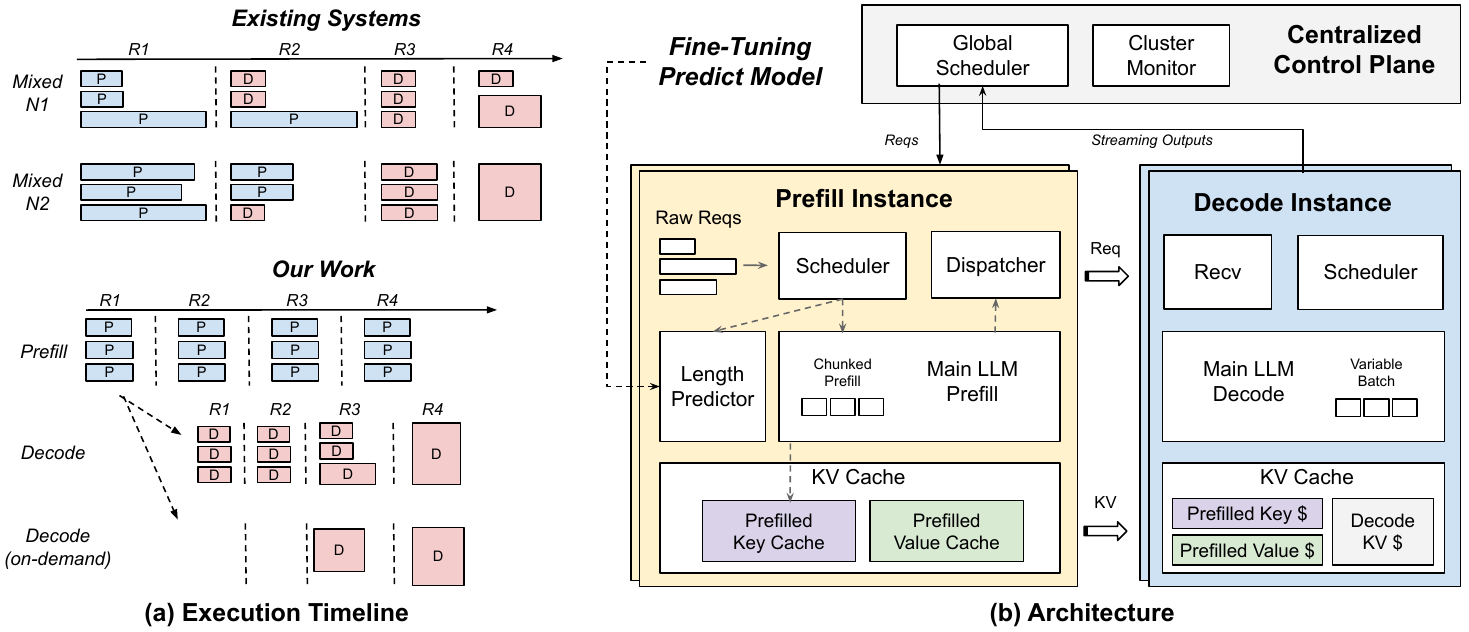}}
\caption{\textbf{\sysname's Workflow and Architecture.}
(a) compares existing systems and \sysname's execution timeline. The existing systems part has two nodes running mixed prefill and decode. \sysname\ has separated prefill and decode instances. This allows for the load-balancing of long decoding tasks across different on-demand decode instances. Each timeline comprises four rounds (R1 to R4), with the length of prefill and decode boxes representing their sequence length and the width of the decode box indicating its resource usage. A wider decode box indicates the presence of lengthy generated tokens, resulting in larger resource usage and decoding latency.
(b) shows \sysname's architecture with four core modules highlighted.}
\label{fig-architecture}
\end{center}
\end{figure*}
}
\section{Design}

\subsection{Overview}

We realize the above insights in \sysname, an LLM inference serving system designed to battle interferences.
First, we run prefill in a fixed-size computation unit by partition and pad input prompts into fixed-size chunks such that the accelerator always runs close to its computation-saturated limit (\S\ref{sec-prefill-instance}).
Second, we design instances dedicated to running the prefill or decode phases. We schedule prefill requests to prefill instances only, and the same goes for decode requests. Prefill instances will transfer prefilled KV cache to decode instances.
Our prefill and decode instances are virtual concepts in that each can scale independently and flip roles if load changes (\S\ref{sec-instance-transition}).
Finally, we design a two-level scheduling algorithm for both prefill and decode request scheduling. We incorporate a length-prediction model to speculate decode requests' resource usage and then schedule them accordingly (\S\ref{sec-decode-instance}).

We show \sysname's architecture in Figure~\ref{fig-architecture} (b) with four modules highlighted: centralized control plane, prefill instance, decode instance, and length prediction model.

\textbf{Centralized control plane.}
It consists of a global scheduler and a cluster monitor.
The global scheduler sends requests to prefill instances based on load and receives streaming outputs from decode instances.
The cluster monitor collects statistics from prefill and decode instances and regularly broadcasts load information to prefill instances. It adds, removes, and flips prefill or decodes instances.

\textbf{Prefill Instances.}
They only run the prefill phase of an LLM inference request.
Each prefill instance has a local scheduler, a length predictor, the main LLM engine, and a dispatcher.
All requests undergo four steps.
First, the local prefill scheduler sorts requests based on pre-defined policies.
Second, the length predictor runs a prediction model to speculate the requests' decode lengths, which are then used to estimate resource usage during the decoding phase.
Third, the main LLM engine partitions all requests into fixed chunks.
Finally, for each request, the dispatcher runs an inter-decode load-balancing algorithm to select a decode instance and then forwards the generated KV cache to it.

\textbf{Decode instances.} 
They are virtually disaggregated from prefill instances and only run the decode phase of an LLM inference request.
Each decode instance can receive requests from any prefill instance.
It runs a local scheduler with three pre-defined policies for selecting decode requests to run in the main LLM engine.

\textbf{Length Prediction Model.}
The prediction model is a small LLM model fine-tuned offline for predicting the generation length of LLM inference requests. \sysname's prefill dispatcher and decode instance's local scheduler utilize the speculated information to schedule decode instances and avoid hotspots measured in \S\ref{sec-bg-dd}. The prediction model is small and deployed at all prefill instances.

\subsection{Control Plane}
\label{sec-control-plane}

\sysname\ has a centralized control plane to
manage inference clusters at the cloud scale.
It consists of a cluster monitor that manages \textit{the lifecycle of prefill and decode instances} and a global scheduler that manages\textit{the lifecycle of inference requests}.
The centralized control plane is a distributed system without a single point of failure or processing bottlenecks.

The cluster monitor is responsible for collecting and broadcasting statistics and scaling instances. Both prefill and decode instances regularly send their load information to the cluster monitor (e.g., every 100 ms). Since we run decentralized decode request scheduling at prefill instances, the cluster monitor will aggregate decode instances' load information and broadcast it to all prefill instances. 

The global scheduler is responsible for forwarding inference requests from external services to prefill instances and sending inference outputs from decode instances back to external services in a streaming fashion.
The global scheduler maintains a request status table, which stores requests' arrival time, current phase (e.g., prefill or decode), SLA requirement, etc.
When a request arrives, the global scheduler will choose a prefill instance with the least load and then insert the request into the table.
%
Following our insight to disaggregate prefill and decode instances, the global scheduler only decides which prefill instance will handle the request. It is up to the prefill instance's dispatcher to decide which decode instances to use with a speculated resource usage.

\subsection{Prefill Instance}
\label{sec-prefill-instance}

The prefill instance runs the prefill phase of an inference request.
To avoid interference among prefill requests, we use a prefill scheduler and chunked prefill to sort and partition all prompts into fixed-size chunks.
To help avoid interference during the decode phase, we run a length predictor and a decentralized dispatcher to choose decode instances based on speculated resource usage.

\subsubsection{Prefill Scheduler}
\label{sec-prefill-scheduler}

The prefill instance's scheduler is crucial for improving the prefill phase's latency and throughput.
The scheduler maintains a raw request queue that stores requests from the global scheduler and a scheduled queue that stores sorted requests.
In this work, we have designed and implemented three scheduler policies: first-come-first-serve (\textit{FCFS}), shortest-job-first (\textit{SJF}), and longest-job-first (\textit{LJF}).
We can use the latter two policies because we can accurately estimate a request's prefill time based on the number of tokens in its prompt.
We only explore non-preemptive policies, though chunked prefill (described soon) has opened the door to preemptive and out-of-order prefill scheduling, such as shortest-remaining-time-first, which we leave for future work.

The scheduled requests are sent to the length predictor  which executes scheduled requests as-is using fixed-size batch (\S\ref{sec-length-predictor}), and the main LLM which uses chunked prefill (\S\ref{sec-chunked-prefill}).
In Figure~\ref{fig-prefill-scheduler}, we illustrate the above three scheduler policies and how scheduled requests are partitioned and merged into fixed-size chunks.
Specifically, FCFS keeps the original request arrival order.
Prompt tokens are partitioned and merged into chunks sequentially.
This policy is the easiest to implement and works best for inference requests with similar prompt lengths.
However, FCFS can lead to head-of-line blocking and high average job completion time (JCT) when requests have long prompts. This is problematic since the length differences among LLM inference requests are more than three orders of magnitude (see Figure~\ref{fig-dataset-cdf}).

In response, we add the shortest-job-first
(SJF), and longest-job-first (LJF) to overcome these issues.
These two policies schedule prefill requests based on prompt token lengths in ascending or descending order. By design, they can achieve lower JCT compared to FCFS. Nevertheless, they are no panacea. They introduce starvation for either long or short requests. To avoid starvation, we propose using a prefill scheduling batch (i.e., \texttt{PrefillSchedBatch}) variable to control how many inference requests can be scheduled at a time. For example, assume the raw request queue has twenty requests awaiting scheduling. If we set the batch size to ten, we will schedule twice, each with ten requests sorted and put into the scheduled queue. This simple mechanism prevents starvation during the prefill phase.

Our scheduler is effective. Results in Figure~\ref{fig-prefill-scheduler-compare} show that SJF lowers average prefill waiting time by 7.8\% compared to FCFS when the batch size is set to 16. Additionaly, the improvement is even more pronounced with larger batch sizes.

{
\begin{figure}[t]
\begin{center}
\centerline{\includegraphics[width=0.4\textwidth]{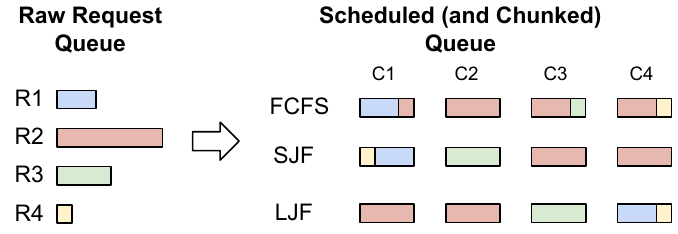}}
\caption{\textbf{Prefill Scheduler Policies.} The left shows four raw inference requests (R1 to R4). The right shows scheduled requests using FCFS, SJF, and LJF. We show the chunked version to illustrate slicing and merging (C1 to C4).}
\vspace{-0.3in}
\label{fig-prefill-scheduler}
\end{center}
\end{figure}
}

\subsubsection{Length Predictor}
\label{sec-length-predictor}

To address the interference cases measured in \S\ref{sec-bg-dd}, it is essential to determine the number of tokens that a decode request is likely to generate. This information will enable us to schedule decode requests in a length-aware manner.
As such, the prefill instance runs a length predictor to predict the length range of an inference request's generated tokens.
The prefill instance's dispatcher utilizes this information for \textit{inter-decode instance} scheduling (\S\ref{sec-dispatcher}), while the decoding instance's local scheduler employs this information for \textit{intra-decode instance} scheduling (\S\ref{sec-decode-instance}).

Our length predictor uses a small LLM-based classification model called a "predict model" to classify the length of generated tokens into fixed-size buckets if the request were executed by a specific target LLM model.
The predict model is intentionally small, containing millions of parameters while the target model is much larger, with billions of parameters. As we run the length predictor at the prefill instance, we aim to minimize its cost and avoid impacting the main LLM model. Therefore, approaches like using a giant LLM to predict length are not feasible for us~\cite{ResponsePredict-arxiv23}.
Fortunately, a small LLM model is much faster than a giant LLM and uses much less resources.
For example, we use OPT-125M as the predict model and OPT-13B as the target model, the small one is roughly ten times faster than the larger one.

We opt to predict the length range instead of an exact number of tokens because the latter is extremely difficult to predict. 
Various inference parameters, such as temperature and top-p~\cite{url-llm-parameter}, result in significant response variations from the same LLM model to the same question in practice. Since our primary goal is to use the estimated length to guide our request scheduling decisions, an exact length estimation is unnecessary; a length range suffices.
For instance, if we estimate the length to be between ten to twenty tokens, we can deduce its resource usage's lower and upper bounds.

In this work, we have tested two execution modes: a sequential mode, where we first execute the predict model followed by the target model, and a parallel mode, where both models are run simultaneously.
The sequential mode adds extra latency for the target LLM model, while the parallel mode may reduce the target LLM model's throughput.
Based on our findings in Figure~\ref{fig-length-prediction-perf}, we opted to use the parallel mode because the main LLM is not affected for most requests (more than 80\%), though throughput take a 10\% hit under extreme stress test. 

Figure~\ref{fig-sft} outlines the offline fine-tuning and online prediction workflow. In this process, the predict model (depicted in red) is trained to speculate the decoding behavior of a specific target model (depicted in blue).
The fine-tuning of the predict model involves three key steps.
Firstly, we assemble a prompt-only training dataset inherited from public datasets, a large target LLM model (e.g., OPT-13B), and a classification model for our predict model (e.g., 125M OPTForSequenceClassification~\cite{url-hg-opt125}). 
Secondly, we send training prompts to the target LLM model, which generates responses.
Subsequently, we categorize the generated responses into fixed-size buckets with a chosen granularity.
For instance, using a granularity of 100, responses with token lengths between 0 to 200 are labeled with 0, 200-400 are labeled with 1, and so on. These labels are paired with the training prompts to create a new dataset. Lastly, we partition the new dataset into a training section and an evaluation section and then proceed to train and evaluate the predict model using this dataset.

The length range granularity plays a crucial role. If set to one, we fall back to predicting an exact number of tokens, which is not practical. If set to target model's context window size (e.g., 2K), we fall back to no prediction at all and could run into interferences reported in \S\ref{sec-bg-pp}. Intuitively, a smaller granularity means more accurate resource and performance estimation but lower accuracy in practice. A larger granularity means higher accuracy but essentially makes scheduling harder. Regardless of granularity, it's easy to calculate resource usage's upper and lower bound but not performance.
In this work, we can predict a granularity of 200 tokens with 74.9\% accuracy.
Since improving prediction accuracy is not the focus of this work, we leave it for future work.

{
\begin{figure}[t]
\begin{center}
\centerline{\includegraphics[width=0.42\textwidth]{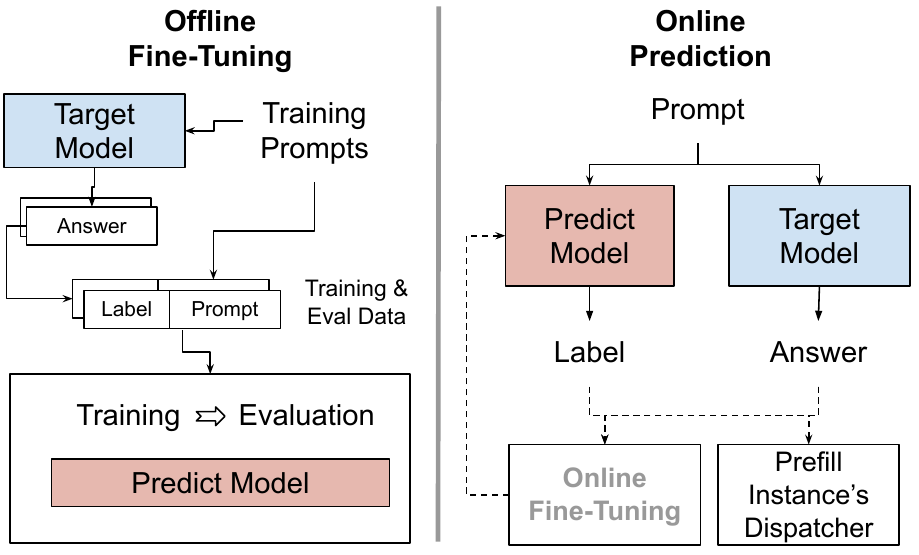}}
\caption{\textbf{Predict Model's Fine-tuning and Prediction Flow.} The target model is the one that we want to predict its decoding behavior. The predict model is the one we train. This work does not explore online fine-tuning.}
\vspace{-0.3in}
\label{fig-sft}
\end{center}
\end{figure}
}

\textbf{Discussions.} 
We run the length predictor at each prefill instance, hence prefill instances can make well-informed decisions on which decode instances should have enough resources to run certain decoding requests.
Nevertheless, we identify two alternative designs. The first design is to run the length predictor at each decode instance. As a result, the prefill instance can only schedule requests based on the load of decoding instances. However, this design cannot avoid interference cases we measured in \S\ref{sec-bg-dd}. Indeed, one could migrate interference requests among decoding instances at runtime based on predicted length. This would be an overly complex solution. The second design is to run the length predictor at the global scheduler before dispatching requests to refill instances. This design could make the global scheduler a bottleneck. We believe our current design is easier and simpler to reason about and deploy compared to alternatives.

\subsubsection{Chunked Prefill}
\label{sec-chunked-prefill}

After the prefill scheduler, we concurrently execute the prefill phase of the main LLM alongside the length predictor.
We employ fixed-size chunks for the LLM prefill rather than using fixed batch sizes~\cite{vllm-sosp23}.

As demonstrated in \S\ref{sec-bg-pp}, we observe that as the number of tokens in a prefill iteration increases, the accelerator's throughput remains constant, while the latency continues to rise after reaching a certain threshold. We refer to this threshold as \texttt{ChunkSize}. Compared to the traditional fixed batch size approach, running prefill in \texttt{ChunkSize} allows for the optimal utilization of accelerators without incurring additional latency penalties. The accelerator and the LLM model architecture determine the \texttt{ChunkSize}. Models with larger hidden dimensions and accelerators with lower capabilities typically result in a smaller \texttt{ChunkSize}. For example, in our test environment, the value is 512 tokens for OPT 13B.

Figure~\ref{fig-prefill-scheduler} illustrates how chunked prefill works for different scheduler policies. For scheduled requests, we first slice and then merge prompt tokens into fixed-size chunks without altering their order. The final chunk in a batch could be partial, and we will pad it to \texttt{ChunkSize} with zeros.
Then, we invoke the main LLM model to execute prefill forward one chunk at a time.
To record progress, we maintain a simple variable per request that records the last prefilled token position.

The benefits of using chunked prefill and various prefill scheduler policies are substantial. In Figure~\ref{fig-prefill-scheduler-compare}, we compare vanilla vLLM, which uses fixed batch size for prefill, against \sysname\, which uses chunked prefill along with FCFS, SJF, and LJF. Chunked prefill with FCFS lowers average prefill latency by 86.4\%. Additionally, we avoid the interference cases measured in \S\ref{sec-bg-pp} as \HP\ requests are broken into fixed-chunks and the accelerator is best utilized.

\textbf{Discussion.}
(1) An early work, Sarathi~\cite{sarathi-arvix23}, has also proposed chunked prefill for the same purpose, where they utilize prefill-decode-mixed chunks. In contrast, our approach involves running prefill-only chunks as we disaggregate LLM's prefill and decode into separate instances.
(2) Our length predictor utilizes a small LLM model for prediction and continues using fixed-size batching instead of chunked prefill. This is due to the model's small size, which does not exhibit a clear compute-saturate threshold as seen in larger models.

\subsubsection{Dispatcher}
\label{sec-dispatcher}

The prefill instance's final module is the dispatcher, which carries out two essential steps for each prefilled request. First, it runs an inter-decode instance scheduling algorithm to select a decode instance and then transmits the prefilled KV cache to the chosen instance.
The dispatcher runs on an event-driven basis, running whenever there are prefilled requests (or chunks).  The disaptcher plays a vital role in mitigating decode and decode interferences as measured in \S\ref{sec-bg-dd}.

Once a request's initial chunk is prefilled, the dispatcher invokes a decentralized load-balancing algorithm to select a decode instance with sufficient resources to run this request's decode phase. 
Our algorithm consists of three steps.
First, we categorize decode instances into two sets: $\alpha$, those with enough resources to execute the chosen request, and $\beta$, those without.
Recall that the prefill instance has all decode instances' load information broadcasted from the cluster monitor (\S\ref{sec-control-plane}). With the predicted length range (\S\ref{sec-length-predictor}), finding decode instances with adequate resources for executing this request's decode phase is easy.
Second, we use the power-of-two~\cite{url-power-of-two} algorithm to choose two instances from the $\alpha$ set randomly. 
Lastly, from the two instances, we choose the one that would encounter the least interference if the prefilled request is sent to it. Based on Figure~\ref{fig-study-decode-decode}, our goal is to establish the lowest average ratio of \HD:\LD, which means we need to spread \HD\ requests evenly. Figure~\ref{fig-decode-strategy-fcfs} proves our algorithm is effective, achieving the lowest total decoding time compared to other policies.

Once a decision is made, the dispatcher sends this request's metadata and prefilled KV cache to the selected decode instance.
Crucially, there are two key design considerations for transferring KV Cache: (1) transfer granularity and (2) network stack for transferring the cache.

We begin by discussing granularity.
Due to our use of chunked prefill, the prefilled KV cache is created in chunks  (\S\ref{sec-chunked-prefill}).
As a result, we have the option to transfer the KV cache either at a  \textit{chunk-level}, sending each chunk's KV cache as it is generated, or at a \textit{request-level}, sending the aggregated KV cache for a request until all of its chunks are prefilled. Utilizing chunk-level transfer enables us to parallelize chunked prefill and KV cache transfer, while request-level transfer allows us to minimize the number of network transfers by sending larger data.
A concurrent work~\cite{patel2023splitwise} has proposed layer-wise KV cache transfer, which aligns with our chunk-level approach. Combining their layer-wise approach with our chunk-level transfer could further optimize compute and network parallelization.
In this work, we only implement request-level transfer for simplicity and leave the chunk-level transfer to future work.

{
\begin{figure}[t]
\begin{center}
\centerline{\includegraphics[width=0.38\textwidth]{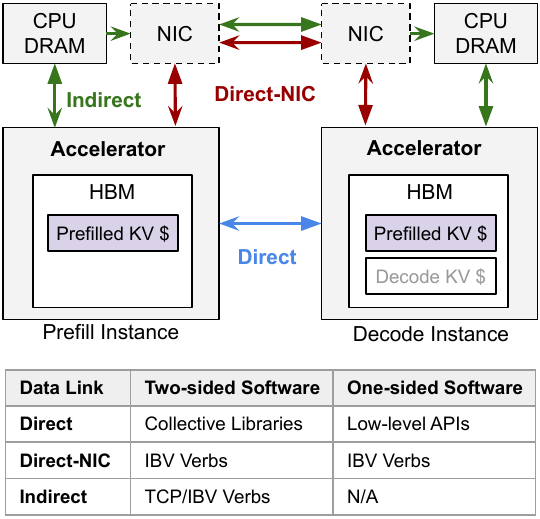}}
\caption {\textbf{Data Links and Network Stacks.} The \textit{Direct} link provides bandwidth in the hundreds of GBs (e.g, NVLink 900GBps). The \textit{Direct-NIC} and \textit{Indirect} link offer bandwidth in the hundreds of Gbs (e.g, ConnectX-6 200Gbps). The 1-sided stack means the sender accelerator can transmit data to the receiver accelerator without involving the receiver's CPU.}
\vspace{-0.2in}
\label{fig-network-stack}
\end{center}
\end{figure}
}

We now delve into the network stack.
Once the main LLM completes its prefill phase, the prefilled KV cache is generated at the accelerator's memory (e.g., GPU's HBM).
Our goal is to transmit this cache to the selected decode instance' accelerator memory, regardless of the hardware platforms on which our system is deployed.
This is challenging as multiple physical data links exist between the prefill and decode instances, each requiring different software stacks.
We classify existing physical data links into three types, as shown in Figure~\ref{fig-network-stack}.
The first is called \textbf{Direct}, where accelerators have a directly connected high-speed link such as NVLink~\cite{url-nvlink} or HCCS~\cite{url-hccs}.
We can use low-level memory copy primitives~\cite{url-cuda-memcpy,url-cann-memcpy} or collective libraries~\cite{url-nccl} to transmit data over these links.
The second is called \textbf{Direct-NIC}, in which accelerators communicate via their companion NICs. We can use custom-built libraries~\cite{url-gpudirect} to transmit data over PCIe \textit{and} Ethernet (or Infiniband).
The third is called \textbf{Indirect}, where there is no direct link, and the accelerators must bounce data via their companion CPU DRAM, incurring extra memory copies.
In Figure~\ref{fig-network-stack}, we also categorize network stacks that utilize aforementioned data links into \textbf{one-sided} and \textbf{two-sided}, similar to RDMA's classification.
Accelerators like GPU or NPU can do one-sided memory access as they have low-level primitives such as direct memory copies between devices~\cite{url-cuda-memcpy,url-cann-memcpy}.

To navigate the complicated physical data links and ensure that \sysname\ can always use the most performant link once deployed, we design a unified network transfer abstraction to utilize the different network stack options listed in Figure~\ref{fig-network-stack}. The stack exposes APIs such as send, receive, read, write, etc. Our dispatcher calls these APIs to transmit the KV cache to remote decode instances.

\textbf{Discussion.}
We identify two unexplored research questions.
The first question pertains to whether it is beneficial to simultaneously utilize multiple data links for transmitting the KV cache. While this approach could enhance performance, it may also introduce complex control logic.
The second question involves the sender accelerator accessing the memory of the receiver accelerator without involving the receiver's CPU. This scenario raises typical challenges associated with building large-scale RDMA-based memory systems~\cite{farm-nsdi12,clio-asplos22}.
Unfortunately, we cannot explore either of these ideas in this work due to limited access to high-end hardware.

\subsection{Decode Instance}
\label{sec-decode-instance}

The decode instance runs the decoding phase of an inference request.
As shown in Figure~\ref{fig-architecture},
it includes a receiver module, which is part of the unified network transfer module, a local scheduler, and an LLM for decoding.
The processing steps are straightforward.
The receiver module accepts requests transmitted from remote prefill instances and waits for prefilled KV caches to be received before adding them to the local scheduler's queue.
The scheduler uses continuous batching to group dynamic-sized batches and invokes the LLM for decoding in an auto-regressive fashion.
We implement \sysname\ based on vLLM~\cite{vllm-sosp23} (see \S\ref{sec-implementation}). Hence, it manages the KV cache in pages rather than reserved for the maximum context length~\cite{fastertransformer-21, yu2022orca}.

With the predicted length information sent from the prefill instance, we propose two \textit{working-set-aware} scheduling policies in addition to vLLM's vanilla one.
vLLM's existing policy schedule requests in a \textbf{greedy} fashion. As long as the accelerator has spare memory, it will add requests to the current iteration. However, it may run out of memory in future iterations and cause thrashing. Fundamentally, it is oblivious to the working set size.
 
To address this limitation, we propose \textbf{reserve-static} and \textbf{reserve-dynamic} policies, both aim to prevent triggering swaps.
Under the reserve-static policy, a request is scheduled only if its predicted memory usage is smaller than the available accelerator memory for the current iteration.
In contrast, the reserve-dynamic policy takes a more proactive approach by considering the predicted number of remaining tokens.
Specifically, a new request is added to the scheduled batch only if there is still spare memory when the shortest remaining job in the batch finishes. This approach effectively mitigates memory thrashing while maximizing the advantages of paging.
Our tests in Figure~\ref{fig-decode-strategy-intra-node} suggest that with our current prediction accuracy, these two policies are on par with vLLM's greedy policy. When the prediction accuracy increases, these two policies can lower the average JCT by roughly 10\%.

\subsection{Instance Flip}
\label{sec-instance-transition}

{
\begin{figure}[t]
\begin{center}
\centerline{\includegraphics[width=0.45\textwidth]{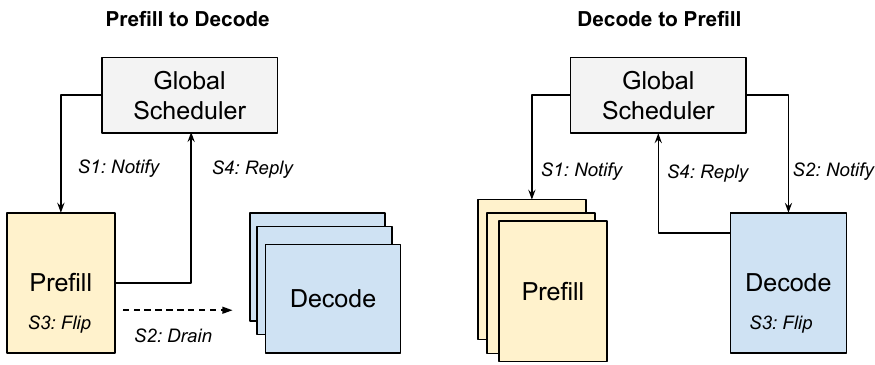}}
\caption{\textbf{Prefill and Decode Instance Flip.}}
\vspace{-0.3in}
\label{fig-instance-transition}
\end{center}
\end{figure}
}

\sysname\ scales out by allocating more hardware resources. Additionally, \sysname\ can also dynamically adjust the number of prefill and decode instances within fixed hardware resources.
This is crucial as LLM inference workloads have huge variations regarding prefill and decode needs (see Figure~\ref{fig-dataset-cdf}), and we cannot statically provision the ratio of prefill and decode instances in advance. Below, we will describe our policy and mechanism to flip a prefill instance to become a decode instance and vice versa.

\textbf{Policy.} As we described in \S\ref{sec-control-plane}, the centralized control plane oversees all instances and has the latest load information. We design a transition watcher module that regularly checks load and decides whether certain instances should be flipped. Various policies can be plugged in, such as flipping an instance if its load has been under 10\% for the past minute.

\textbf{Mechainsm.}
Once an instance is selected, we will go through the steps depicted in Figure~\ref{fig-instance-transition}.
To flip a prefill instance, the global scheduler stops forwarding requests and then sends a flip request to it. The prefill instance will wait until all queued requests are drained. Then, we flip the instance. 
Flipping a decode instance is slightly more complex. The global scheduler notifies all prefill instances to stop forwarding requests to the selected decode instance and notifies the decode instance to complete flipping.
Note that the actual flipping is fast and simple. It involves changing an internal variable without restarting the process or reloading models. In our current implementation, both instance flips take roughly 5 to 7 ms, excluding the dynamic draining time.

\section{Implementation}
\label{sec-implementation}

We implement \sysname's centralized control plane from scratch in Python. We adopt prefill and decode instances based on vLLM~\cite{vllm-sosp23}.
Most of our core modules are implemented in Python, except for the unified network stack, which utilizes C++ to interface with low-level APIs and IB Verbs for network transfer. Additionally, we implement a shared-memory-based communication mechanism that enables fast command transfer across Python and C++ languages. The fine-tuning part uses Trainer APIs offered by HuggingFace Transformer~\cite{url-hg-opt125}.

A prefill or decode instance is a single deployable unit consisting of two processes when deployed.
For prefill, it has a Python process that runs the scheduler, length predictor, and the main LLM, as well as a C++ process that runs the dispatcher and the network stack.
For decode, it has a Python process for running the scheduler and the main LLM, along with a C++ process that runs the network stack.

Due to limited high-end hardware availability, our current implementation only supports the \textit{Indirect} type using sockets (see Figure~\ref{fig-network-stack}). 
In order to evaluate \sysname's performance across different hardware configurations, we have implemented a mock mechanism to emulate varying network bandwidth. This mechanism works as follows: for a given set of requests, we initially run their prefill phase offline to obtain their prefilled KV cache. Before testing, we load these prefilled KV caches into the decode instance's local memory. When testing starts, the prefill instance transmits only the request metadata to the decode instance, excluding the actual prefilled KV cache. Subsequently, the decode instance calculates the latency of the KV cache transfer and waits accordingly. This latency is calculated given a specific model architecture and the hardware bandwidth we aim to emulate.

\section{Evaluation}
\label{sec-eval}

We evaluate \sysname\ using public dataset~\cite{sharegpt} and report time-to-first-token (TTFT), job completion time (JCT), and efficiency as in performance per dollar (perf/\$).

Our testbed consists of four NVIDIA V100 GPUs, each with 32GB HBM.
All GPUs are plugged into a single server with Xeon Gold 5218R CPU and 256GB DRAM.
For the large LLM, we run OPT-13B~\cite{opt-arxiv22}.
For the length prediction model, we use OPT-125M.
We compare with vLLM~\cite{vllm-sosp23}.
Since we adopted \sysname\ after vLLM, both systems manage KV caches in pages. 
Unlike \sysname, vanilla vLLM tightly couples prefill and decode phases.

\subsection{End-to-End Performance}
\label{sec-app-cost-perf}


This section compares \sysname\ with vanilla vLLM using end-to-end benchmarks.
We emulate \sysname\ atop two hardware setups using the mock mechanism described in \S\ref{sec-implementation}. The first is \textit{TS-RoCE}, assuming prefill and decode instances communicate over 200Gbps RoCE (Direct-NIC in Figure~\ref{fig-network-stack}). The second is called \textit{TS-NVLink}, assuming instances communicate over 300GBps NVLink (Direct in Figure~\ref{fig-network-stack}). Both setups are adopted from commercial V100-based servers.

For all tests, the prefill instance's scheduler uses the SJF policy as it has the best performance with the \texttt{PrefillSchedBatch} set to 16 (see Figure~\ref{fig-prefill-scheduler-compare}).
For the inter-decode instance scheduling, we use the decentralized load-balancing algorithm as it outperforms other policies.
For the intra-decode instance scheduling, we use the reserve-dynamic policy atop paging as it could outperform vLLM's greedy policy (see Figure~\ref{fig-decode-strategy-intra-node}).
We flip an instance once it becomes idle for a minute using mechanisms described in \S\ref{sec-instance-transition}.

To understand how these systems perform under mixed downstream inference tasks,
we run five different types of workloads as presented in Figure~\ref{fig-dataset-cdf}:
Heavy Prefill with Light Decode (\textit{HPLD}),
Heavy Prefill with Heavy Decode (\textit{HPHD}),
Light Prefill with Heavy Decode (\textit{LPHD}),
Light Prefill with Light Decode (\textit{LPLD}),
and \textit{Mixed}.
Akin to the configuration used in \S\ref{sec-bg-motivation}, prefill requests that have more than 512 prompt tokens are categorized as heavy, and others are light. Decode requests with more than 128 tokens are categorized as heavy as ShareGPT answers' median length is 128.
We generate these workloads using samples from the ShareGPT dataset~\cite{sharegpt}, following the distribution illustrated in Figure~\ref{fig-dataset-cdf}.

For each workload, we compare all systems across three key metrics: \textit{TTFT}, \textit{JCT}, and \textit{resource usage time} (i.e., cost).
Comparing TTFT indicates whether \sysname's prefill scheduler and chunked prefill are effective.
Comparing JCT indicates whether \sysname's disaggregated prefill and decode design and two-level decode scheduling are effective.
A natural question that centers around our disaggregated design is cost.
Intuitively, \sysname\ uses two times the resources compared to vLLM's prefill-decode-coupled setting.
Our results suggest otherwise. Two factors contributed: first, \sysname's prefill and decode run faster; second, \sysname\ can recycle or flip instances to reduce waste.
Below, resource usage time represents the aggregated wall time that the prefill and decode instances use to run a particular workload. For example, the resource usage time is 3 seconds if we run prefill in in 1 second and decode in 2 seconds. For vLLM, it is the total runtime since it couples prefill and decode.

We now present each workload's results.

\textbf{Light Prefill and Light Decode.}
Generally, LPLD represents the chat workload.
We  test LPLD in Figure~\ref{fig-ben-lpld-dataset} using 128 requests.
When comparing \sysname\ to vLLM, we reduce average TTFT by 44\% and average JCT by 40\% for both emulated hardware setups.
Despite using twice the number of hardware cards, \sysname\ completes tasks almost twice as fast, resulting in resource usage time that is comparable to the vanilla vLLM. Thus, we improve perf/\$ by 1.4x.

\textbf{Light Prefill and Heavy Decode.}
Generally, LPHD represents the content creation workload.
We test LPHD in Figure~\ref{fig-ben-lphd-dataset} using 128 requests.
Surprisingly, \sysname\ improves average TTFT by 97\% despite using short prompts.
This is because vLLM's prefill incurs serious interference while running prefill and decode requests in the same batch; in contrast, \sysname\ disaggregates them into separate instances.
Additionally, with variable decode batch size over vLLM's fixed batch size during the decode phase, \sysname\ improves average JCT by 47\% while using 38\% less total hardware resources. Overall, we improve perf/\$ by 2.4x.

\textbf{Heavy Prefill and Light Decode \& Heavy Prefill and Heavy Decode}.
HPLD and HPHD represent summarization or prompt engineering types of workloads.
Both have long prompt tokens.
This means \sysname\ faces two challenges: (a) large prefilled KV caches and (b) the main LLM may be impacted by the prediction model (roughly 10\% as shown in Figure~\ref{fig-length-prediction-perf}).
Nevertheless, in Figure~\ref{fig-ben-hpld-dataset}, we can see that \sysname\ still improves average TTFT and average JCT by 9\% and 23\%, respectively, but at the cost of 43\% increase in resource usage.
vLLM outperforms \sysname\ in terms of perf/\$ by 14\%.
As Figure~\ref{fig-ben-hphd-dataset} shows, with heavy decode, \sysname's TTFT improvement is more pronounced because we disaggregated heavy decode from prefill, akin to Figure~\ref{fig-ben-lphd-dataset}. We improve the average JCT by 19\% at the cost of 7\% more resources, improving perf/\$ by 1.1x.

\textbf{Mixed.} 
The last workload is a mix of all the above workloads, randomly sampled from the ShareGPT dataset.
This is the case where a cluster is running all kinds of requests.
In Figure~\ref{fig-ben-mix-dataset}, we run 128 requests, \sysname\ lowers average TTFT, average JCT, and resource usage by 85\%, 50\%, 21\%, respectively, improving perf/\$ by 1.9x.



\textbf{Takeaways.}
(1) For most LLM inference workloads, \sysname\ improve average TTFT, average JCT, resource use time, and most importantly, perf/\$ by a large margin.
(2) Disaggregating prefill from decode into two distinct instances significantly improves TTFT and efficiency by minimizing interference, particularly for workloads with heavy decodes such as LPHD and HPHD.
(3) \sysname's design is not ideal for HPHD workloads as the room for improvement is small, and the overhead we introduce cannot be offset.


{
\begin{figure*}[ht]
    \vspace{5pt}
    \setlength{\abovecaptionskip}{3pt plus 1.0pt minus 1.0pt}
    \setlength{\belowcaptionskip}{6pt plus 1.0pt minus 1.0pt}
  \begin{minipage}[t]{0.49\linewidth}
    \includegraphics[width=\linewidth]{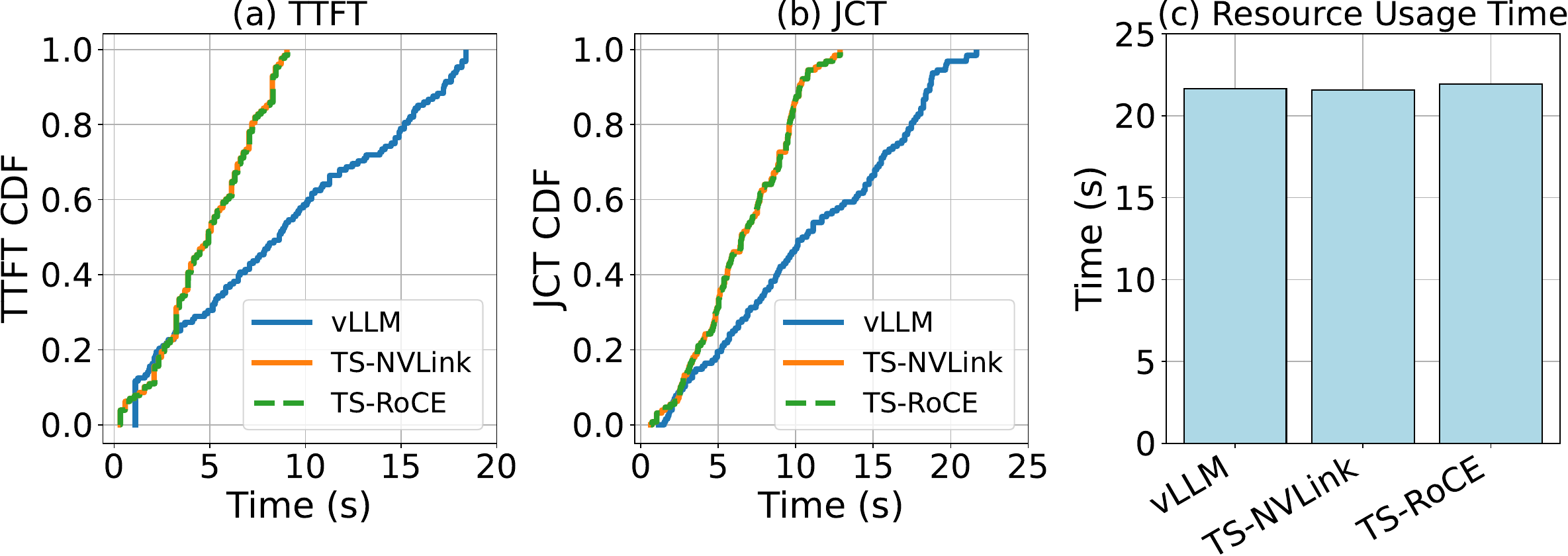}
    \caption{\textbf{Light Prefill and Light Decode (LPLD)}}
    \label{fig-ben-lpld-dataset}
  \end{minipage}%
    \hspace{0.01\linewidth}
  \begin{minipage}[t]{0.49\linewidth}
        \includegraphics[width=\linewidth]{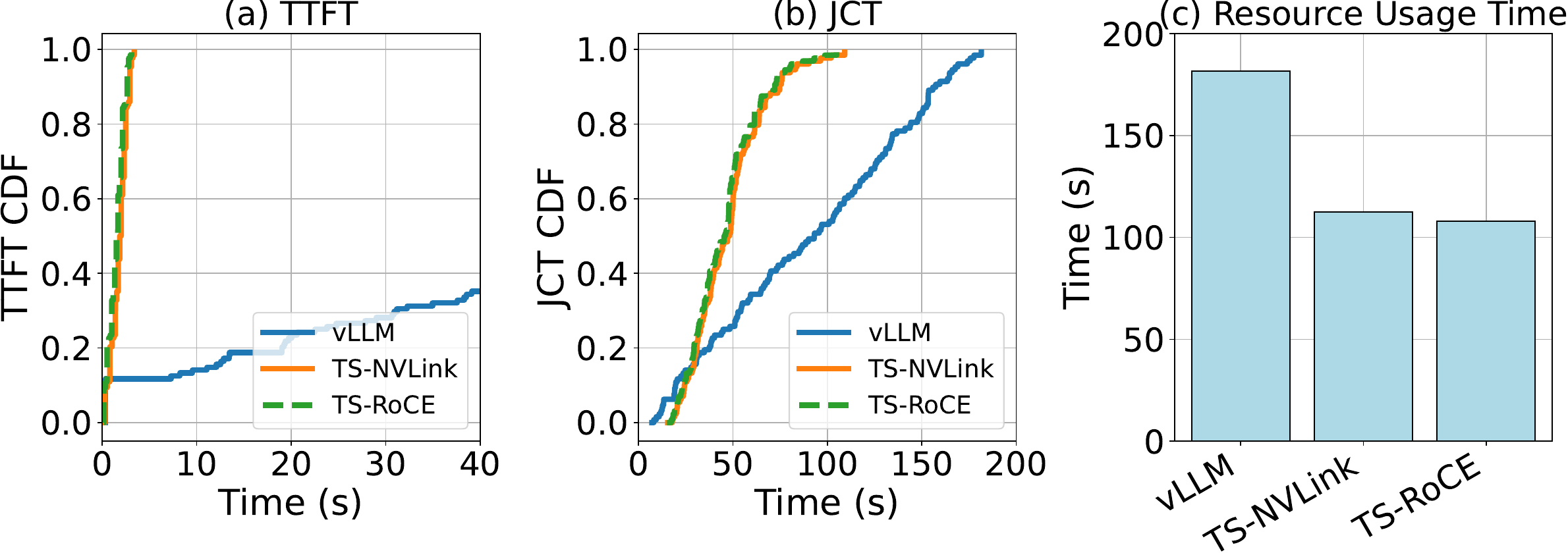}
        \caption{\textbf{Light Prefill and Heavy Decode (LPHD)}}
        \label{fig-ben-lphd-dataset}
  \end{minipage}
  \begin{minipage}[t]{0.49\linewidth}
        \includegraphics[width=\linewidth]{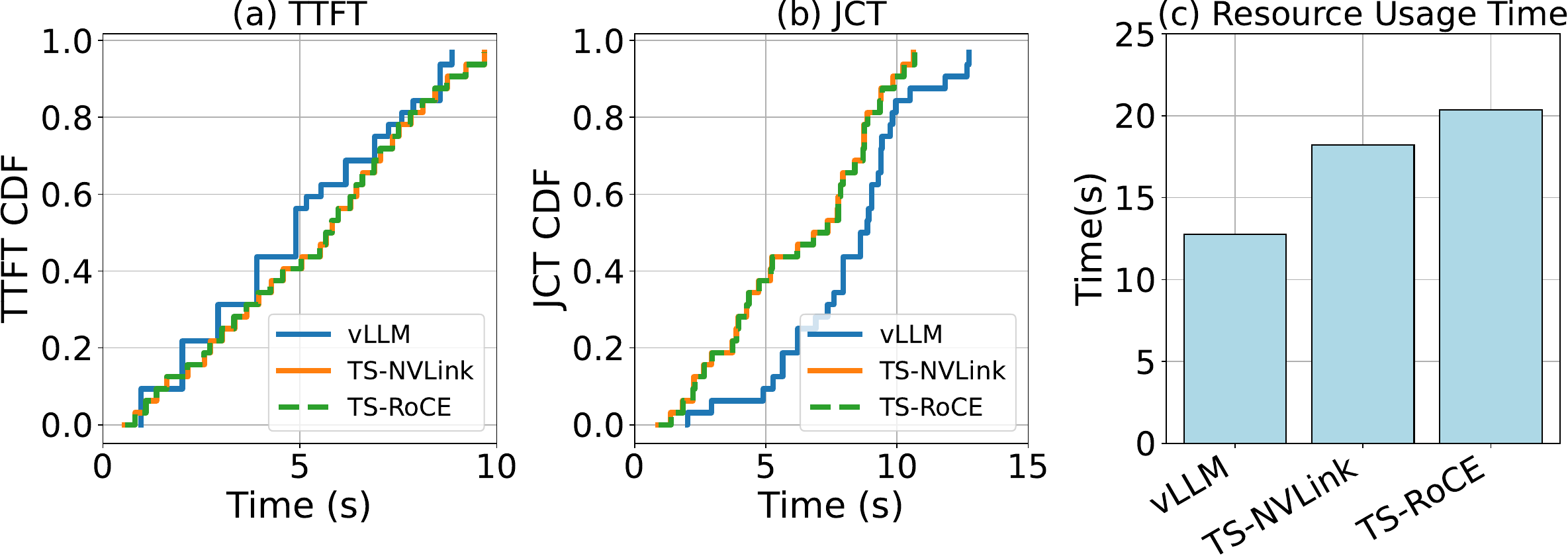}
        \caption{\textbf{Heavy Prefill and Light Decode (HPLD)}}
        \label{fig-ben-hpld-dataset}
  \end{minipage}%
    \hspace{0.01\linewidth}
  \begin{minipage}[t]{0.49\linewidth}
    \includegraphics[width=\linewidth]{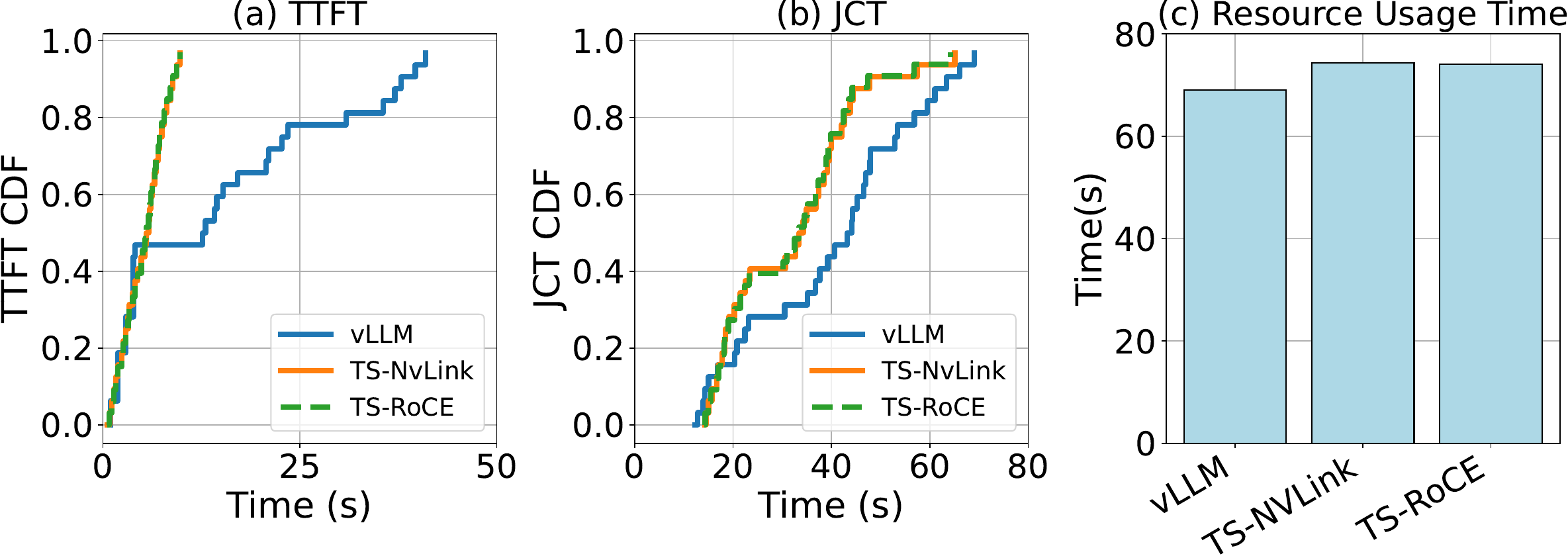}
    \caption{\textbf{Heavy Prefill and Heavy Decode (HPHD)}}
    \label{fig-ben-hphd-dataset}
  \end{minipage}
  \begin{minipage}[t]{0.49\linewidth}
    \includegraphics[width=\linewidth]{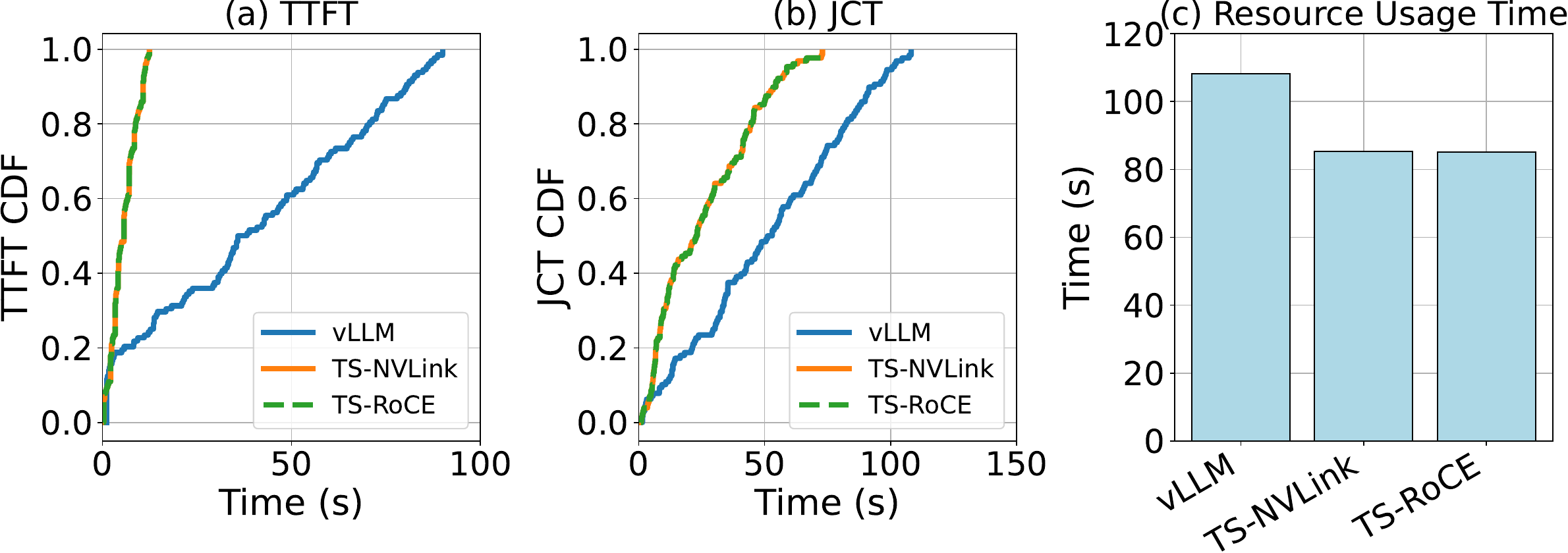}
    \caption{\textbf{Mixed Prefill and Decode}}
    \label{fig-ben-mix-dataset}
  \end{minipage}%
  \hspace{0.01\linewidth}
  \begin{minipage}[t]{0.49\linewidth}
    \includegraphics[width=\linewidth]{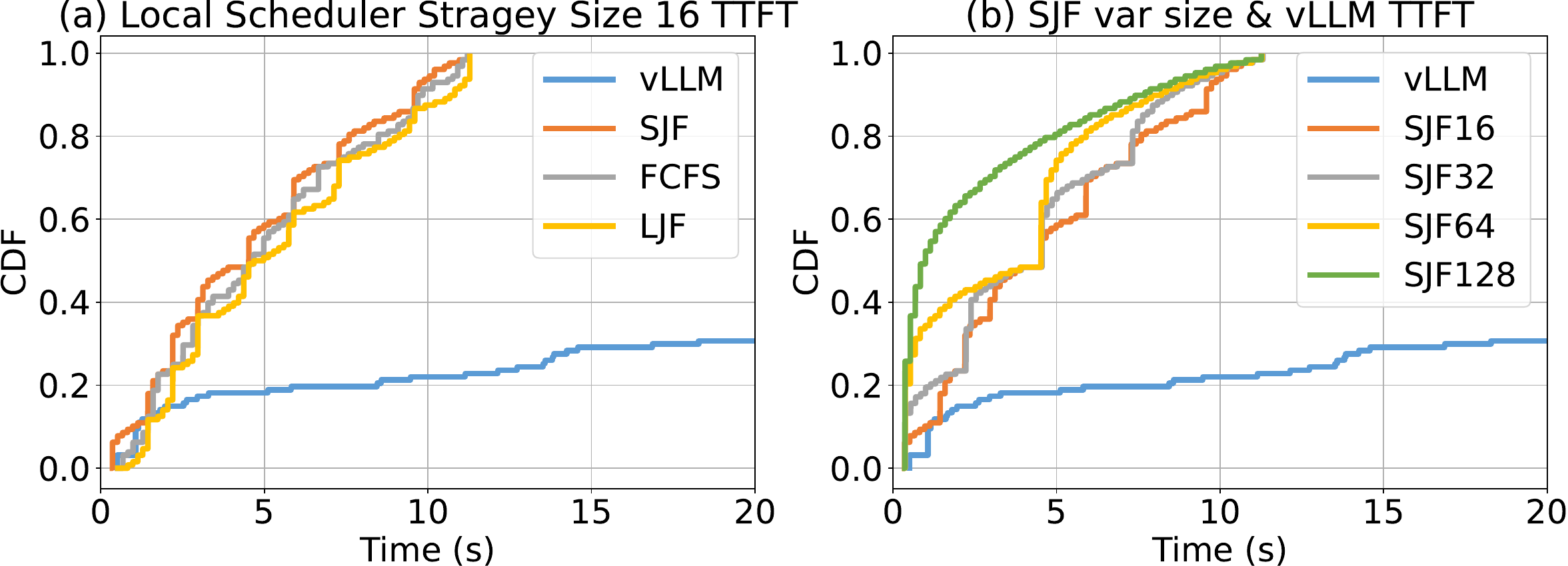}
    \caption{\textbf{Scheduler Policies and Chunked Prefill.}} 
    \label{fig-prefill-scheduler-compare}
  \end{minipage}
  \label{fig:main}
\end{figure*}
}

\subsection{Microbenchmark}
\label{sec-eval-microbench}

\subsubsection{Prefill Scheduler}
\label{sec-prefill-scheduler-cost-perf}


Below, we study the overhead of sorting requests, compare different policies, and study the impact of batch size on performance.
Note we run OPT-13B TP=2, \texttt{ChunkSize} is set to 512, and vLLM's batch size is set to 16.
 
\textbf{Sort.}
Our scheduler sorts incoming requests based on the length of their input tokens if non-FCFS policies are used.
We use Python's native sort API.
We find the sorting overhead ranges from 10s to 100s of microseconds,
which is negligible compared to millisecond-level or second-level TTFT latency.

\textbf{Scheduler Policy and Batch Size.}
In Figure~\ref{fig-prefill-scheduler-compare},
we compare \sysname\, which uses chunked prefill along with FCFS, SJF, and LJF,
against vanilla vLLM, which uses fixed batch size for prefill.
Requests used in this test follow the ShareGPT distribution.
In the left part, we set \texttt{PrefillSchedBatch} to 16.
Compared to vLLM's fixed batch mode, chunked prefill alone with FCFS improves latency by 86.4\%. Additionally, the SJF scheduler policy further lowers the average waiting time by 7.8\%.
The right part examines the impact of adjusting \texttt{PrefillSchedBatch}. When we increase the batch size from 16 to 128, SJF's average TTFT decreases by 46.5\%.
The improvement in TTFT increases with a larger scheduling batch.

\subsubsection{Length Predictor}
\label{sec-length-predictor-cost-perf}

Our length predictor uses the OPT-125M classification model to speculate the decoding behavior of the OPT-13B model  (\S\ref{sec-length-predictor}).
This section studies the performance of both models and the prediction accuracy.

In Figure~\ref{fig-length-prediction-perf},
we run stress tests among several settings regarding per-iteration latency and throughput.
\textit{L-Alone} means running the OPT-13B model alone, using chunked fill with \texttt{ChunkSize} set to 512.
\textit{P-Alone} means running the OPT-125M prediction model alone. It does not use chunked prefill but uses dynamic batch sizes. It can group multiple requests into a batch. Due to the limitation of ~\cite{url-hg-opt125}, we need to pad requests in a batch to the longest one. For example, if we have two requests in a batch, one has 100 tokens, and the other has 500 tokens. Then, we need to pad the first to 500 tokens. This is costly for requests with short prompts. Hence, we set a cutting limit. Requests higher than the limit will run alone. The default is 512 tokens.
\textit{L+P512} means OPT-13B model's performance if co-runs with the small OPT-125M in parallel. The suffix number means the max padded size.
We can see that the large LLM's prefill latency is roughly ten times of the small LLM's.
If we co-run both models and a padding limit of 512, 80\% of large LLM's prefill requests remain unchanged compared to when it runs alone. 
Overall, while co-running with a small LLM, the large LLM's average prefill latency increases by 10\%, and throughput drops by 12\%. Note that these are stress tests. The impact will be smaller in practice.
We believe beefier hardware can further mitigate the drop.


We train our OPT-125M prediction model using 75K training data from ShareGPT. 
We test the model using three different length range granularities: 100, 200, and 400. The accuracy achieved by our prediction model for these granularities is 58.9\%, 74.9\%, and 85\%, respectively.


\subsubsection{Decode Scheduling}
\label{sec-decode-scheduling-cost-perf}

{
\begin{figure}[t]
\begin{center}
\vspace{5pt}
\centerline{\includegraphics[width=0.48\textwidth]{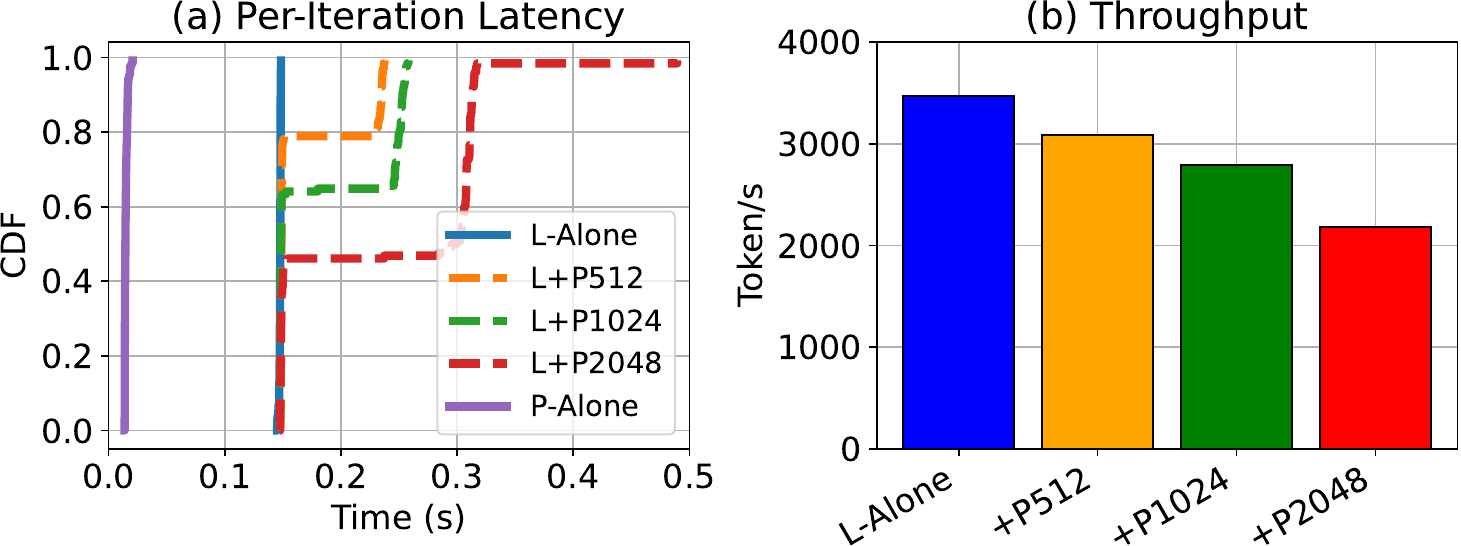}}
\caption{\textbf{Running Large LLM with Prediction Model.}}
\label{fig-length-prediction-perf}
\end{center}
\end{figure}
}
{
\begin{figure}[t]
\begin{center}
\centerline{\includegraphics[width=0.48\textwidth]{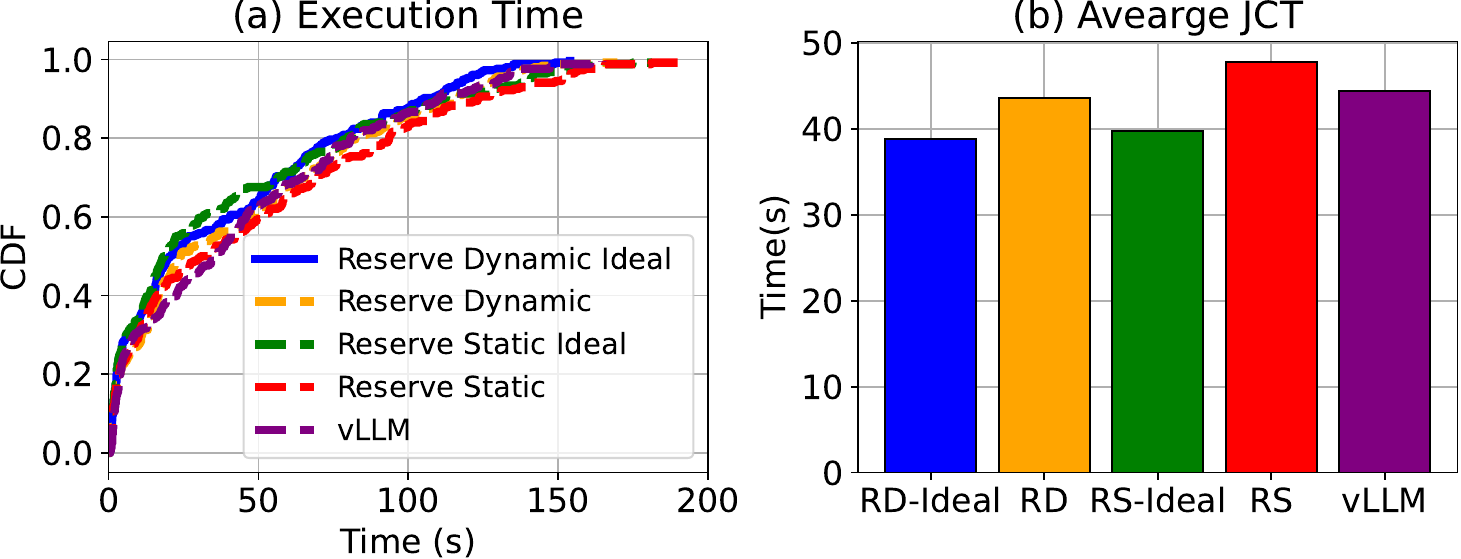}}
\caption{\textbf{Intra-Decode Instance Scheduling.} }
\label{fig-decode-strategy-intra-node}
\end{center}
\end{figure}
}
{
\begin{figure}[t]
\begin{center}
\vspace{5pt}
\centerline{\includegraphics[width=0.48\textwidth]{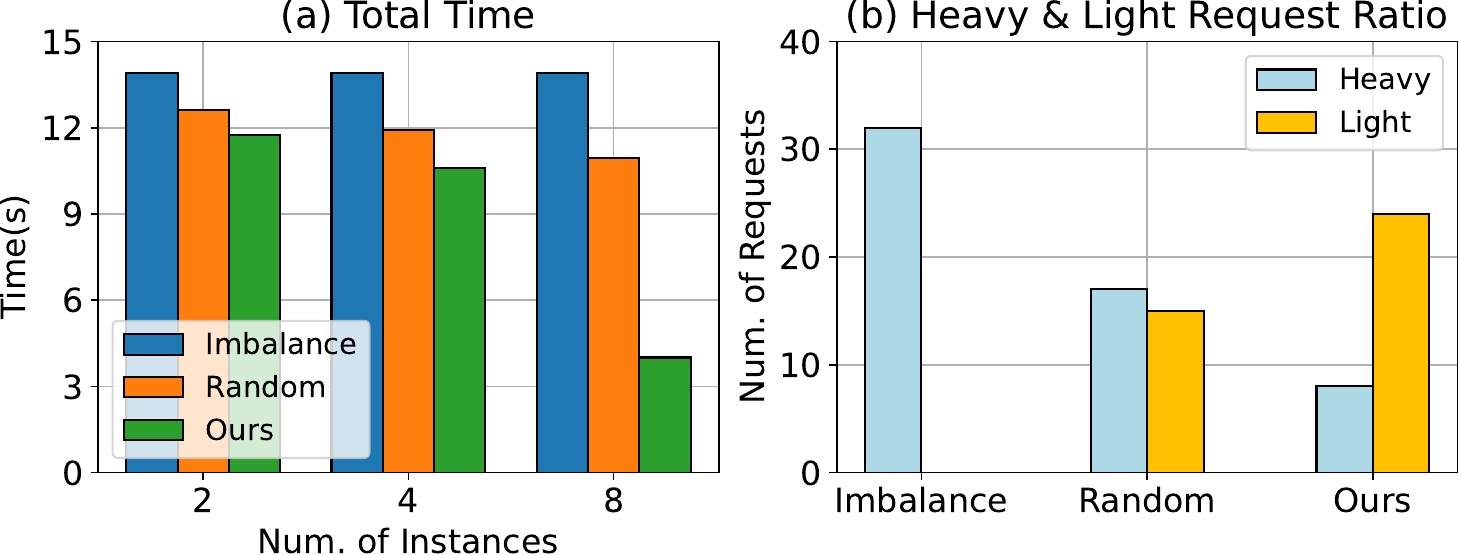}}
\caption{\textbf{Inter-Decode Instance Scheduling.} }
\label{fig-decode-strategy-fcfs}
\end{center}
\end{figure}
}

We now study the scheduling policies related to decode instances. We first compare intra-decode instance scheduler policies (\S\ref{sec-decode-instance}) and then compare different load balance algorithms for inter-decode instance scheduling (\S\ref{sec-dispatcher}). All tests use OPT-13B with TP=2.

We compare three intra-decode scheduler algorithms, namely vLLM's greedy, \sysname's reserve-static (RS), and reserve-dynamic (RD) in Figure~\ref{fig-decode-strategy-intra-node}.
We run 256 requests following ShareGPT distribution.
Our policies estimate resource usage using the predicted length range's lower end. We compare using the actual accuracy (acc-200 74.9\%) and an ideal accuracy of 100\%.
While using the actual accuracy,
reserve-dynamic achieves the same JCT as vLLM's greedy algorithm. When using an ideal prediction accuracy, 
reserve-dynamic and reserve-static improve average JCT by 12\% and 10\%, respectively. This is because our policies carefully provision requests based on their memory usage.

We compare three distributed load balance algorithms in Figure~\ref{fig-decode-strategy-fcfs}.
Firstly, we present our \textit{decentralized power-of-two} algorithm, designed to distribute requests based on predicted length. 
The second is \textit{random}, in which the prefill instance randomly chooses a decode instance.
The third algorithm, \textit{imbalance}, simulates a worst-case scenario where heavy decode requests are consistently directed to the same decode instances. We run 32 requests per decode instance, spanning the range of 2 to 8 decode instances.
Figure~\ref{fig-decode-strategy-fcfs}'s left part shows that \sysname's decentralized load balancing algorithm is effective, achieving the lowest total decoding time. The right parts show the number of \HD\ and \LD\ requests in the slowest instance. Cleary, \sysname's inter-decode scheduling algorithm evenly balances load across instances, which avoids interferences measured in ~\S\ref{sec-bg-dd}.

\section{Related Work}

\begin{table}\footnotesize
    \centering
    \vspace{5pt}
    \begin{tabular}{c c  c  c  c }
    \hline
       \scriptsize\bf Work & \scriptsize\bf C. P. & \scriptsize\bf Disagg. P/D & \scriptsize\bf Interference &  \scriptsize\bf Dist-Sched. \\
    \hline
    \hline
       \bf \sysname \  &  \checkmark & \checkmark & \checkmark &  \checkmark \\
       Splitwise\cite{patel2023splitwise} \  &  \scalebox{0.85}{$\times$} &  \checkmark &  \scalebox{0.85}{$\times$} & \checkmark \\
       Sarathi\cite{sarathi-arvix23} \  &  \checkmark &  \scalebox{0.85}{$\times$} &  \scalebox{0.85}{$\times$} & \scalebox{0.85}{$\times$}\\
       vLLM\cite{vllm-sosp23} \ & \scalebox{0.85}{$\times$} & \scalebox{0.85}{$\times$} & \scalebox{0.85}{$\times$} & \scalebox{0.85}{$\times$} \\
       FastServe\cite{fastserve-arxiv23} \  & \scalebox{0.85}{$\times$} & \scalebox{0.85}{$\times$} & \scalebox{0.85}{$\times$} & \checkmark \\
    \hline
    \end{tabular}
    \caption{\textbf{Related work comparison.}
    (1) C. P.: chunked prefill.
    (2) Disagg. P/D: disaggregated prefill and decode. 
    (3) Interference: whether the system deals with inference interference.
    (4) Dist-Sched: distributed scheduling policies.
    } 
    \label{table-related-work}
\end{table}

Table~\ref{table-related-work} compares \sysname\ with other closely related works.
We are among the first to disaggregate prefill and decode in LLM inference, concurrent to Splitwise~\cite{patel2023splitwise}.
Sarathi\cite{sarathi-arvix23} has proposed chunked prefill to overcome suboptimal prefill processing. They run prefill-decode-mixed chunks. In contrast, \sysname\ runs prefill-only chunks as we observe non-neglible interference between prefill and decode, thus choose to disaggregate prefill from decode.
FastServe\cite{fastserve-arxiv23} utilizes a multi-level priority feedback queue to minimize JCT. In contrast, \sysname\ utilizes two-level scheduling for prefill and decode instances. Our policies are working-set-aware, reducing interference and swaps thereby improving JCT and efficiency.
%
%
%
Many recent work focus on optimizing batching, caching, and scheduling~\cite{deepspeed-sc22, alpaserve-arvix23, tritonserver-21}.
Specifically, Orca~\cite{yu2022orca} introduce the iterative-level scheduling. Sheng et.al~\cite{sheng2023fairness} have proposed a fair scheduler based on the continuous batching mechanism.
%
%
%
Many works try to optimize memory usage. For example, using quantization\cite{rethinking-icml20,smoothquant-icml23,Llmint8-arvix23, flexgen-pmlr23, PTQ-arixv23, gptzip-icml23, spqr-arvix22, optq-eiclr22} to compress the model weights into lower precision, using paging to reduce fragmentation~\cite{vllm-sosp23}, and low-level algorithm and kernel optimizations~\cite{zeroquant-nips22, rammer-osdi20, lightseq-arvix20, flashattention-nips22, flashattention2-arxiv23, flashdecoding++-2023, bytetransformer-ipdps23}.
Those works are orthogonal to our efforts to mitigate interference.

\section{Conclusion}

We propose \sysname, an LLM inference serving system designed to battle interference.
Our key insight is to carefully schedule and group inference requests based on their characteristics.
It has three key parts. First, it partitions prompts into fixed-size chunks, ensuring the accelerator consistently operates at its computation-saturated limit. Second, it disaggregates prefill and decode instances to avoid interference when mixing them together. Finally, it uses a smart two-level scheduling algorithm to avoid decode scheduling hotspots. Results show that \sysname\ improves time-to-first-token, job completion time, and inference efficiency in terms of performance per dollar by a large margin.


\bibliographystyle{plain}
\bibliography{paper}

\end{document}